\documentclass[12pt]{article}%
\usepackage{amssymb}
\usepackage{amsmath}
\usepackage{amsfonts}
\usepackage{graphicx}%
\providecommand{\U}[1]{\protect\rule{.1in}{.1in}}

\begin{document}
\bigskip%
\begin{titlepage}
\vspace{.3cm} \vspace{1cm}
\begin{center}
\baselineskip=16pt \centerline{\bf{\Large
{ Resolving Cosmological Singularities}}}\vspace{2truecm}
\centerline{\large\bf Ali H.
Chamseddine$^{1,2}$\ , Viatcheslav Mukhanov$^{3,4,5}$\ \ } \vspace{.5truecm}
\emph{\centerline{$^{1}%
$Physics Department, American University of Beirut, Lebanon}}
\emph{\centerline{$^{2}$I.H.E.S. F-91440 Bures-sur-Yvette, France}}
\emph{\centerline{$^{3}%
$Niels Bohr International Academy, Niels Bohr Institute, Blegdamsvej 17, DK-2100 Copenhagen, Denmark}%
}
\emph{\centerline{$^{4}%
$Theoretical Physics, Ludwig Maxmillians University,Theresienstr. 37, 80333 Munich, Germany }%
}
\emph{\centerline{$^{5}%
$MPI for Physics, Foehringer Ring, 6, 80850, Munich, Germany}}
\end{center}
\vspace{2cm}
\begin{center}
{\bf Abstract}
\end{center}
We find a simple modification of the longitudinal mode in General Relativity which incorporates  the idea of limiting
curvature. In this case the singularities in contracting Friedmann and Kasner universes are avoided, and instead,
the universe has a regular bounce which takes place during the time inversely proportional to the square root
of the limiting curvature. Away from the bounce, corrections to General Relativity are negligible. In addition
the non-singluar modification of General Relativity delivers for free a realistic candidate for Dark Matter.
\end{titlepage}%

\section{Introduction}

One of the long standing unsolved problem in General Relativity is the problem
of spacelike singularities. In fact, assuming that General Relativity is
universally valid and imposing rather general conditions on the state of
matter, Hawking and Penrose have proven that space-times describing such as,
for instance, Friedmann and Kasner universes and black holes, are geodesically
incomplete $\cite{PenHawk}$. Gravitational collapse leads to singularities
where the curvature invariants grow and become infinite. It is well known that
when the curvature approaches the Planckian value quantum effects become
extremely important and Einstein theory must be modified. This is the reason
for the common belief that a nonperturbative quantum gravity must finally
resolve the singularities. In fact, near the singularity, vacuum polarization
induces higher order curvature corrections to the Einstein action and particle
production becomes very important \cite{GH}. These effects must completely
modify General Relativity when the Planck curvature is approached.
Unfortunately, there is no reliable way to calculate constructively all these
effects nonperturbatively and therefore the \textquotedblleft quantum
resolution\textquotedblright\ of singularities remains rather obscure, and at
the moment it is not more than wishful thinking. In fact progress along this
line of research has been relatively modest in spite of the enormous efforts
devoted to the problem.

There is a different \textquotedblleft non-quantum\textquotedblright\ approach
to resolving the singularities. In fact, one can imagine that at high
curvatures, classical General Relativity is modified to incorporate the idea
of limiting density/curvature \cite{Markov}. If this limiting curvature is at
least few orders of magnitude below the Planckian value, then one can safely
ignore nonperturbative quantum gravity effects and can entirely rely on the
classical theory used in investigating the singularities. On the other hand,
if the limiting curvature is large enough and the theory reduces to General
Relativity at low curvature, it will not be in contradiction with experiment
and therefore such a theory will be fully legitimate. Once again the progress
in solving the singularity problem using this approach is rather modest. In
the papers \cite{MukBran} \cite{Mukbransol} it was suggested to modify
Einstein's theory by adding some complicated combination of higher order
curvature invariants which allowed obtaining a non-singluar homogeneous
isotropic universe. However, it is totally unclear how to extend the proposal
in these papers, which is rather messy and complicated, to the cases of
anisotropic singularities and the singularity \textquotedblleft
inside\textquotedblright\ the black hole.

In this letter we shall consider a minimal simple modification of Einstein's
theory, where the singularities are avoided at the classical level in
Friedmann and Kasner \cite{Landau} universes irrespective of the matter which
fills the universe. In an accompanying paper we will show that black holes in
this theory are also nonsingluar. This opens the possibility to have a theory
of gravity where singularities are absent in general. To construct such a
theory we will introduce the constrained scalar field $\phi,$ which satisfies
\begin{equation}
g^{\mu\nu}\partial_{\mu}\phi\partial_{\nu}\phi=1\label{0}%
\end{equation}
As it was shown in \cite{mimetic}, this constrained scalar field is not
dynamical by itself, but induces mimetic dark matter in Einstein theory making
the longitudinal degree of freedom of the gravitational field dynamical.
Notice that its emergence can, for example, be justified in noncommutative
geometries as a consequence of quantization of thee-dimensional volume
\cite{Quanta}. In a previous paper \cite{mimcos}, by allowing a potential term
$V\left(  \phi\right)  ,$ we were able to find bouncing nonsingular solutions
for isotropic, homogeneous universe, but even in this case, the most general
solutions remain singular. In addition to this drawback, the term $V\left(
\phi\right)  $ explicitly spoils time translational invariance because in the
synchronous coordinate system the generic solution of (\ref{0}) is $\phi
=t+$\textrm{constant. }We have also found that the potential term does not
help in avoiding the singularity in Kasner universe and inside the black hole.
Therefore we will assume here the absence of a potential term by requiring
invariance with respect to the shift symmetry $\phi\rightarrow\phi
+$\textrm{constant, }and instead add to the Einstein action a function
$f\left(  \square\phi\right)  $, which is invariant\textrm{.} The appearance
of such function $f$ can be easily justified in the spectral action approach
of noncommutative geometry \cite{Scale} \cite{Uncanny}. Clearly we cannot
derive it in nonperturbative way, but we can try just to find a theory where
this function allows to resolve singularities in General Relativity. In
particular, we will show that there exist Born-Infeld type of actions with
\[
f\left(  \square\phi\right)  =1-\sqrt{1-\frac{\left(  \square\phi\right)
^{2}}{\varepsilon_{m}}}+....,
\]
and for which singularities are resolved, that is, the contracting universes
bounce at the limiting curvature and all curvature invariants always remains
regular and bounded by the values, characterized by $\varepsilon_{m}.$ Once
again we would like to stress that introducing $\square\phi$ does not add to
the system any new dynamical scalar fields and new degrees of freedom. The
\textquotedblleft field $\phi$\textquotedblright\ always remains constrained
by (\ref{0}) and in the synchronous coordinate system it just
\textquotedblleft serves\textquotedblright\ as time, making the longitudinal
degree of freedom of the gravitational field to be dynamical. Thus, this
theory must be viewed as a modification of General Relativity in the
longitudinal sector. Because the longitudinal gravitational field induced by
matter via the constraint has a \textquotedblleft negative
energy\textquotedblright\ by itself, it is not surprising that the proposed
modification of Einstein's theory violates the conditions needed in the proof
of singularity theorems and hence these singularities can be avoided. Although
noncommutative geometry offers a very strong support for the model considered
here \cite{Scale}, we will not require, or use any information or methods from
that framework. Instead our purpose is to propose nonsingular classical
modification of General Relativity irrespective of its justification from the
point of view of so called \textquotedblleft fundamental
theory\textquotedblright, which is not known at present.

\section{The theory}

Let us consider the theory described by the action
\begin{equation}
S=\int\left(  -\frac{1}{2}R+\lambda\left(  g^{\mu\nu}\partial_{\mu}%
\phi\partial_{\nu}\phi-1\right)  +f\left(  \chi\right)  +L_{m}\right)
\sqrt{-g}d^{4}x, \label{1}%
\end{equation}
where $\chi=\square\phi$, $\lambda$ is a Lagrange multiplier, $L_{m}$ is the
Lagrangian of the usual matter and we set $8\pi G=1$. The mimetic scalar field
$\phi$ satisfies the constraint \cite{mimetic} \cite{mimcos}%
\begin{equation}
g^{\mu\nu}\partial_{\mu}\phi\partial_{\nu}\phi=1, \label{2}%
\end{equation}
and therefore the term $f\left(  \chi\right)  $ does not lead to the
appearance of higher derivatives and ghost degrees of freedom. Variation of
action $\left(  \ref{1}\right)  $ with respect to the metric gives the
following equations%
\begin{equation}
G_{\mu\nu}=R_{\mu\nu}-\frac{1}{2}g_{\mu\nu}R=\tilde{T}_{\mu\nu}+T_{\mu\nu},
\label{3}%
\end{equation}
where $T_{\mu\nu}$ is the energy-momentum tensor of the usual matter and
\begin{equation}
\tilde{T}_{\mu\nu}=2\lambda\partial_{\mu}\phi\partial_{\nu}\phi+g_{\mu\nu
}\left(  \chi f^{\prime}-f+g^{\rho\sigma}\partial_{\rho}f^{\prime}%
\partial_{\sigma}\phi\right)  -\left(  \partial_{\mu}f^{\prime}\partial_{\nu
}\phi+\partial_{\nu}f^{\prime}\partial_{\mu}\phi\right)  , \label{4}%
\end{equation}
with $f^{\prime}=df/d\chi,$ describes the extra contribution to Einstein
equations due to the $\phi$ dependent terms. In the synchronous coordinate
system the metric takes the form \cite{Landau}%
\begin{equation}
ds^{2}=dt^{2}-\gamma_{ik}\left(  t,x^{l}\right)  dx^{i}dx^{k}, \label{5}%
\end{equation}
and if no coordinate singularities arise, the most general solution of
constraint equation $\left(  \ref{2}\right)  $ is \cite{Landau}
\begin{equation}
\phi=\pm t+A, \label{6}%
\end{equation}
where $A$ is a constant of integration. In this coordinate system, the
coordinate independent invariant $\chi=\square\phi$ becomes%

\begin{equation}
\chi=\square\phi=\frac{1}{\sqrt{-g}}\frac{\partial}{\partial x^{\mu}}\left(
\sqrt{-g}g^{\mu\nu}\frac{\partial\phi}{\partial x^{\nu}}\right)  =\frac
{\dot{\gamma}}{2\gamma}, \label{7}%
\end{equation}
with $\gamma=\det\gamma_{ik}$ and dot denotes time derivative. In this paper
we will consider flat Friedmann and Kasner universes, where the metric
$\left(  \ref{5}\right)  $ depends only on time $t,$ that is, $\gamma
_{ik}=\gamma_{ik}\left(  t\right)  .$ In this case the nonvanishing components
of the curvature are \cite{Landau}%
\begin{equation}
R_{0}^{0}=-\frac{1}{2}\dot{\varkappa}-\frac{1}{4}\varkappa_{i}^{k}%
\varkappa_{k}^{i},\,\qquad R_{k}^{i}=-\frac{1}{2\sqrt{\gamma}}\frac{d\left(
\sqrt{\gamma}\varkappa_{k}^{i}\right)  }{dt}, \label{8}%
\end{equation}
where $\varkappa_{k}^{i}=\gamma^{im}\dot{\gamma}_{mk}$, $\varkappa
=\varkappa_{i}^{i}=\dot{\gamma}/\gamma$. The components of mimetic matter
contribution to the Einstein's equations are
\begin{align}
\tilde{T}_{0}^{0}  &  =2\lambda+\chi f^{\prime}-f-\dot{\chi}f^{\prime\prime
},\nonumber\\
\tilde{T}_{k}^{i}  &  =\left(  \chi f^{\prime}-f+\dot{\chi}f^{\prime\prime
}\right)  \delta_{k}^{i}, \label{9}%
\end{align}
The equation
\begin{equation}
R_{0}^{0}-\frac{1}{2}R=\tilde{T}_{0}^{0}+T_{0}^{0}%
\end{equation}
then takes the form%
\begin{equation}
\frac{1}{8}\left(  \varkappa^{2}-\varkappa_{i}^{k}\varkappa_{k}^{i}\right)
=2\lambda+\chi f^{\prime}-f-\dot{\chi}f^{\prime\prime}+T_{0}^{0}, \label{10}%
\end{equation}
The space-space equation
\begin{equation}
R_{k}^{i}=\tilde{T}_{k}^{i}-\frac{1}{2}\tilde{T}\delta_{k}^{i}+T_{k}^{i}%
-\frac{1}{2}T\delta_{k}^{i},
\end{equation}
where $\tilde{T}=\tilde{T}_{\alpha}^{\alpha}$ and $T=T_{\alpha}^{\alpha}$,
gives%
\begin{equation}
\frac{1}{2\sqrt{\gamma}}\frac{\partial\left(  \sqrt{\gamma}\varkappa_{k}%
^{i}\right)  }{\partial t}=\left(  \lambda+\chi f^{\prime}-f\right)
\delta_{k}^{i}-T_{k}^{i}+\frac{1}{2}T\delta_{k}^{i}. \label{12}%
\end{equation}
To determine the Lagrange multiplier $\lambda$ it is convenient to integrate
the equation obtained by variation of the action with respect to $\phi:$
\begin{equation}
\frac{1}{\sqrt{\gamma}}\partial_{0}\left(  2\sqrt{\gamma}\lambda\right)
=\square f^{\prime}=\frac{1}{\sqrt{\gamma}}\partial_{0}\left(  \sqrt{\gamma
}f^{\prime\prime}\dot{\chi}\right)  , \label{14}%
\end{equation}
from which it follows that
\begin{equation}
\lambda=\frac{C}{2\sqrt{\gamma}}+\frac{1}{2}f^{\prime\prime}\dot{\chi}.
\label{15}%
\end{equation}
where $C$ is a constant of integration corresponding to mimetic cold matter.

Let us assume that for the usual matter $T_{k}^{i}=-p\delta_{k}^{i}.$ This
assumption is valid for all models we consider below and it is generic enough
to understand how matter can influence singularities in case of general
space-times. In this case, by subtracting from equation $\left(
\ref{12}\right)  $ one third of its trace, we obtain%
\begin{equation}
\frac{\partial}{\partial t}\left(  \sqrt{\gamma}\left(  \varkappa_{k}%
^{i}-\frac{1}{3}\varkappa\delta_{k}^{i}\right)  \right)  =0, \label{16}%
\end{equation}
and therefore
\begin{equation}
\varkappa_{k}^{i}=\frac{1}{3}\varkappa\delta_{k}^{i}+\frac{\lambda_{k}^{i}%
}{\sqrt{\gamma}}, \label{17}%
\end{equation}
where $\lambda_{k}^{i}$ are constants of integration which satisfy
$\lambda_{i}^{i}=0.$ Taking into account that $\varkappa=2\chi=\dot{\gamma
}/\gamma$ and substituting expression $\left(  \ref{17}\right)  $ together
with $\left(  \ref{15}\right)  $ into $\left(  \ref{10}\right)  $ we arrive
at
\begin{equation}
\frac{1}{3}\chi^{2}+f-\chi f^{\prime}=\frac{\lambda_{k}^{i}\lambda_{i}^{k}%
}{8\gamma}+\frac{C}{\sqrt{\gamma}}+T_{0}^{0}, \label{18}%
\end{equation}
which is a first order differential equation for $\gamma$. Solving this
equation and substituting the result in $\left(  \ref{17}\right)  $ we can
determine all components of the metric.

There exists a large class of functions $f$ which lead to singularity free
solutions. We require that curvature invariants must be bounded by some
limiting maximal values determined by $\chi_{m}$ which is smaller than the
Planck value. In this case we can ignore quantum effects and the classical
non-singluar solutions we obtain below, will be completely legitimate
irrespective of quantum corrections, even close to the bounce at limiting
curvatures. It is obvious that at $\chi^{2}\ll\chi_{m}^{2}$ the corrections to
general relativity will be negligible only if the expansion of $f\left(
\chi\right)  $ at small $\chi$ starts at order $\chi^{4}.$ To demonstrate this
idea and obtain non-singluar solutions we will take the function $f\left(
\chi\right)  $ to be of the Born-Infeld type\footnote{The use of the
multivalued functions $\arcsin\chi$ and $\sqrt{1-\chi^{2}}$ could be avoided
by first defining a function $\chi=\sin\psi.$ We thank Alain Connes for
pointing this to us.}%
\begin{equation}
f\left(  \chi\right)  =1+\frac{1}{2}\chi^{2}-\chi\arcsin\chi-\sqrt{1-\chi^{2}}%
\end{equation}
which after scaling $\chi\rightarrow\sqrt{\frac{2}{3}}\frac{\chi}{\chi_{m}}$
and $f\rightarrow\chi_{m}^{2}f$ \ takes the form
\begin{equation}
f\left(  \chi\right)  =\chi_{m}^{2}\left(  1+\frac{1}{3}\frac{\chi^{2}}%
{\chi_{m}^{2}}-\sqrt{\frac{2}{3}}\frac{\chi}{\chi_{m}}\arcsin\left(
\sqrt{\frac{2}{3}}\frac{\chi}{\chi_{m}}\right)  -\sqrt{1-\frac{2}{3}\frac
{\chi^{2}}{\chi_{m}^{2}}}\right)  ,\label{18a}%
\end{equation}
and leads, using $\left(  \ref{18}\right)  ,$ to the particularly simple
equation
\begin{equation}
\chi_{m}^{2}\left(  1-\sqrt{1-\frac{2}{3}\frac{\chi^{2}}{\chi_{m}^{2}}%
}\right)  =\varepsilon,\label{19b}%
\end{equation}
We have denoted by $\varepsilon$ the terms in the right hand side of equation
$\left(  \ref{18}\right)  ,$ which are independent on the time derivative of
metric$,$ that is,%
\begin{equation}
\varepsilon=\frac{\lambda_{k}^{i}\lambda_{i}^{k}}{8\gamma}+\frac{C}%
{\sqrt{\gamma}}+T_{0}^{0}.\label{20a}%
\end{equation}
By squaring equation $\left(  \ref{19b}\right)  $ and recalling that
$\chi=\dot{\gamma}/2\gamma,$ equation $\left(  \ref{20a}\right)  $ can be
rewritten as%
\begin{equation}
\frac{1}{12}\left(  \frac{\dot{\gamma}}{\gamma}\right)  ^{2}=\varepsilon
\left(  1-\frac{\varepsilon}{\varepsilon_{m}}\right)  ,\label{21a}%
\end{equation}
where $\varepsilon_{m}=2\chi_{m}^{2}.$

\section{Nonsingular Friedmann universe}

Consider first, the flat isotropic Friedmann universe with the metric
\[
ds^{2}=dt^{2}-a^{2}\left(  t\right)  \delta_{ik}dx^{i}dx^{k}%
\]
In this case $\gamma=a^{6},$ and $\lambda_{k}^{i}=0.$ Correspondingly,
equation $\left(  \ref{21a}\right)  $ becomes
\begin{equation}
3\left(  \frac{\dot{a}}{a}\right)  ^{2}=\frac{\varepsilon_{m}}{a^{3}}\left(
1-\frac{1}{a^{3}}\right)  , \label{22a}%
\end{equation}
where we have normalized the scale factor $a$ in such a way as to have
$\varepsilon=\varepsilon_{m}$ at $a=1$ for mimetic matter and skipped the
contribution of the other matter, that is, $T_{0}^{0}=0.$ This equation can be
easily integrated and its solution
\begin{equation}
a=\left(  1+\frac{3}{4}\varepsilon_{m}t^{2}\right)  ^{1/3}, \label{23a}%
\end{equation}
describes first the contracting universe which is cold matter dominated for
$t<-1/\sqrt{\varepsilon_{m}},$ where $a\propto t^{2/3}$, then it passes
through the regular bounce during time interval $-1/\sqrt{\varepsilon_{m}%
}<t<1/\sqrt{\varepsilon_{m}}.$ Finally after the bounce for $t>1/\sqrt
{\varepsilon_{m}}$ the universe is expanding as the normal \ dust dominated
Friedmann universe with $a\propto t^{2/3}.$ To see how the usual
hydrodynamical matter will influence the bounce, if it dominates at the
limiting curvature, let us neglect mimetic matter and consider the flat
universe filled by matter with constant equation of state $p=w\varepsilon$.
Equation $\left(  \ref{22a}\right)  $ then becomes%
\begin{equation}
3\left(  \frac{\dot{a}}{a}\right)  ^{2}=\frac{\varepsilon_{m}}{a^{3\left(
1+w\right)  }}\left(  1-\frac{1}{a^{3\left(  1+w\right)  }}\right)  ,
\label{24a}%
\end{equation}
and its solution
\begin{equation}
a=\left(  1+\frac{3}{4}\left(  1+w\right)  ^{2}\varepsilon_{m}t^{2}\right)
^{\frac{1}{3\left(  1+w\right)  }} \label{25a}%
\end{equation}
is regular at the bounce. One can easily see that all solutions for the
contracting universe with an arbitrary spacial curvature are regular near the
bounce and all invariants of the curvature remain bounded. In fact, the
spatial curvature term is proportional to $1/a^{2}$ and it can obviously be
neglected at high curvatures compared to the mimetic dust or usual
hydrodynamical matter with equation of state $w>-1/3.$ In the vicinity of the
bounce, which happens when $\varepsilon\left(  a_{m}\right)  =\varepsilon
_{m},$ we can approximate $\left(  \ref{21a}\right)  $ as
\begin{equation}
3\left(  \frac{\dot{a}}{a}\right)  ^{2}\simeq-\left(  \frac{d\varepsilon}%
{da}\right)  _{m}\left(  a-a_{m}\right)  . \label{26a}%
\end{equation}
Since $\varepsilon$ grows as $a$\ decreases the derivative $d\varepsilon/da$
is negative and hence $a$ must be always larger than or equal to $a_{m}.$ The
solution of the equation above, near the bounce, is given by%
\begin{equation}
a\simeq a_{m}+\left\vert \frac{a_{m}^{2}}{12}\left(  \frac{d\varepsilon}%
{da}\right)  _{m}\right\vert t^{2} \label{27a}%
\end{equation}
which remains regular irrespective of the matter content and the spatial
curvature of the universe. Obviously all curvature invariants are also bounded
and regular.

\section{Avoiding the singularity in Kasner universe}

Now we turn to the case of contracting Kasner universe. The Kasner Universe is
the homogeneous, but anisotropic solution of Einstein's equation in empty
space
\begin{equation}
ds^{2}=dt^{2}-t^{2p_{1}}dx^{2}-t^{2p_{2}}dy^{2}-t^{2p_{3}}dz^{2}, \label{27b}%
\end{equation}
where the constants $p_{i}$ satisfy the conditions \cite{Landau}%
\begin{equation}
p_{1}+p_{2}+p_{3}=1,\text{ \ }p_{1}^{2}+p_{2}^{2}+p_{3}^{2}=1. \label{27c}%
\end{equation}
For $t<0$ this metric describes the contracting universe, while for positive
$t$ an expanding homogeneous anisotropic universe. The scalar curvature and
Ricci tensor square for solution $\left(  \ref{27b}\right)  $ vanish, that is,
$R=0$ and $R_{\alpha\beta}R^{\alpha\beta}=0.$ However, space-time is curved
because the Riemann curvature squared invariant is equal to
\begin{equation}
R_{\alpha\beta\gamma\delta}R^{\alpha\beta\gamma\delta}=-\frac{16}{t^{4}}%
p_{1}p_{2}p_{3}, \label{27d}%
\end{equation}
and becomes infinite at $t=0,$ thus showing that there is a final singularity
in contracting universe and initial singularity in expanding universe. We will
now show how in our theory this singularity is resolved by limiting curvature,
and will obtain the bouncing Kasner solution. To find this solution let us
consider the metric
\[
\gamma_{ik}=\gamma_{(i)}\left(  t\right)  \delta_{ik},
\]
the determinant of which is equal to $\gamma=\gamma_{\left(  1\right)  }%
\gamma_{\left(  2\right)  }\gamma_{\left(  3\right)  }.$ To simplify the
formulae we first consider empty universe without any matter and later on will
show why the presence of matter does not change the main conclusion about
absence of the singularity. In empty universe, $\varepsilon$ defined in
$\left(  \ref{20a}\right)  ,$ is given by
\begin{equation}
\varepsilon=\frac{\lambda_{k}^{i}\lambda_{i}^{k}}{8\gamma}, \label{28a}%
\end{equation}
where $\lambda_{k}^{i}$ do not depend on time and and it is traceless,
$\lambda_{i}^{i}=0.$ Equation $\left(  \ref{21a}\right)  $ then becomes
\begin{equation}
\left(  \frac{\dot{\gamma}}{\gamma}\right)  ^{2}=\frac{3\bar{\lambda}^{2}%
}{2\gamma}\left(  1-\frac{\bar{\lambda}^{2}}{8\varepsilon_{m}\gamma}\right)  ,
\label{29a}%
\end{equation}
where we have denoted $\bar{\lambda}^{2}\equiv\lambda_{k}^{i}\lambda_{i}^{k}.$
This equation can be easily integrated and the result is%
\begin{equation}
\gamma=\frac{\bar{\lambda}^{2}}{8\varepsilon_{m}}\left(  1+3\varepsilon
_{m}t^{2}\right)  \label{30b}%
\end{equation}
Thus, the determinant of the metric remains finite and bounded from below. We
now find the components of the metric. Without loss in generality we can
diagonalize $\lambda_{k}^{i}$ so that its eigenvalues are $\lambda_{\left(
i\right)  }=(\lambda_{1},$ $\lambda_{2},$ $\lambda_{3}).$ In this case,
\begin{equation}
\varkappa_{k}^{i}=\gamma^{im}\dot{\gamma}_{mk}=\frac{\dot{\gamma}_{\left(
i\right)  }}{\gamma_{\left(  i\right)  }}\delta_{k}^{i}, \label{31}%
\end{equation}
and from equation $\left(  \ref{17}\right)  $ it follows that the components
of the metric satisfy
\begin{equation}
\frac{\dot{\gamma}_{\left(  i\right)  }}{\gamma_{\left(  i\right)  }}=\frac
{1}{3}\frac{\dot{\gamma}}{\gamma}+\frac{\lambda_{\left(  i\right)  }}%
{\sqrt{\gamma}}. \label{32}%
\end{equation}
Since $\gamma$ is bounded from below and $\dot{\gamma}/\gamma$ always remains
analytical and finite near the bounce all curvature invariants built out of
$\varkappa_{k}^{i}$ and its derivatives are non-singluar. Integrating equation
$\left(  \ref{32}\right)  $ we obtain
\begin{equation}
\gamma_{\left(  i\right)  }=\gamma^{1/3}\exp\left(  \lambda_{\left(  i\right)
}\int\frac{dt}{\sqrt{\gamma}}\right)  , \label{33}%
\end{equation}
and substituting for $\gamma$ from $\left(  \ref{30b}\right)  $ we find%
\begin{equation}
\gamma_{\left(  i\right)  }=\left(  \frac{\bar{\lambda}^{2}}{8\varepsilon_{m}%
}\left(  1+3\varepsilon_{m}t^{2}\right)  \right)  ^{1/3}\exp\left(
2\sqrt{\frac{2}{3}}\frac{\lambda_{\left(  i\right)  }}{\overline{\lambda}%
}\sinh^{-1}\left(  \sqrt{3\varepsilon_{m}}t\right)  \right)  , \label{34}%
\end{equation}
Taking into account that for $\varepsilon_{m}t^{2}\gg1$
\[
\sinh^{-1}\left(  \sqrt{3\varepsilon_{m}}t\right)  \simeq\pm\ln\left\vert
2\sqrt{3\varepsilon_{m}}t\right\vert ,
\]
where the plus and minus signs are to be taken for $t\gg1/\sqrt{\varepsilon
_{m}}$ and $t\ll-1/\sqrt{\varepsilon_{m}},$ respectively, then this metric
simplifies to
\begin{equation}
\gamma_{\left(  i\right)  }\simeq\left(  \frac{\bar{\lambda}^{2}%
}{32\varepsilon_{m}}\right)  ^{1/3}\left(  12\varepsilon_{m}t^{2}\right)
^{p_{i}^{\pm}}, \label{34a}%
\end{equation}
where%
\begin{equation}
p_{i}^{\pm}=\frac{1}{3}\pm\sqrt{\frac{2}{3}}\frac{\lambda_{\left(  i\right)
}}{\overline{\lambda}}. \label{35}%
\end{equation}
Since $\lambda_{1}+$ $\lambda_{2}+$ $\lambda_{3}=0,$ the $p_{i}^{\pm}$ satisfy
the conditions
\begin{equation}
p_{1}^{\pm}+p_{2}^{\pm}+p_{3}^{\pm}=1,\text{ }\left(  p_{1}^{\pm}\right)
^{2}+\left(  p_{2}^{\pm}\right)  ^{2}+\left(  p_{3}^{\pm}\right)  ^{2}=1,
\label{35a}%
\end{equation}
and we have the familiar Kasner solution $\left(  \ref{27b}\right)  $ at low
curvatures. Only when the curvature becomes of order $\varepsilon_{m},$ the
Kasner contraction changes and for $\varepsilon_{m}t^{2}\ll1$ the metric
\begin{equation}
\gamma_{\left(  i\right)  }\simeq\left(  \frac{\bar{\lambda}^{2}}%
{8\varepsilon_{m}}\right)  ^{1/3}\left(  1+\frac{\lambda_{\left(  i\right)  }%
}{\overline{\lambda}}\sqrt{8\varepsilon_{m}}t\right)  , \label{36}%
\end{equation}
describes the regular bounce and finally an expanding Kasner universe. Notice
that during the bounce the indices characterizing the Kasner universe are
changing. If in a contracting universe they were $p_{i}^{-}$ then after the
bounce they become%
\begin{equation}
p_{i}^{+}=\frac{2}{3}-p_{i}^{-}. \label{36o}%
\end{equation}
In particular, the Kasner Universe with $p_{i}^{-}=\left(  -\frac{1}{3}%
,\frac{2}{3},\frac{2}{3}\right)  $ turns after the bounce to the Kasner
universe with $p_{i}^{+}=\left(  1,0,0\right)  $ for which all curvature
invariants, including the Riemann tensor squared given in $\left(
\ref{27d}\right)  ,$ are equal to zero.$\,\ $\ Hence after the bounce the
Kasner universe evolves to Minkowski space-time. In fact, the coordinate
transformation%
\begin{equation}
T=t\cosh x,\text{ \ }X=t\sinh x, \label{36f}%
\end{equation}
brings the Kasner metric
\begin{equation}
ds^{2}=dt^{2}-t^{2}dx^{2}-dy^{2}-dz^{2}, \label{36y}%
\end{equation}
to the standard Minkowski form%
\begin{equation}
ds^{2}=dT^{2}-dX^{2}-dy^{2}-dz^{2}. \label{36yy}%
\end{equation}
The curvature invariants for the non-singluar solution $\left(  \ref{34}%
\right)  $ are bounded and regular everywhere. In fact, substituting $\left(
\ref{30b}\right)  $ into the expressions
\begin{equation}
R=-\dot{\varkappa}-\frac{1}{3}\varkappa^{2}-\frac{\bar{\lambda}^{2}}{4\gamma},
\label{36ff}%
\end{equation}%
\begin{equation}
R_{\alpha\beta}R^{\alpha\beta}=\frac{1}{3}\dot{\varkappa}^{2}+\frac{1}%
{6}\varkappa^{2}\dot{\varkappa}+\frac{1}{36}\varkappa^{4}+\frac{1}{4\gamma
}\left(  \dot{\varkappa}+\frac{1}{6}\varkappa^{2}\right)  \bar{\lambda}%
^{2}+\frac{1}{16\gamma^{2}}\bar{\lambda}^{4} \label{fff}%
\end{equation}

\begin{equation}
R_{\alpha\beta\gamma\delta}R^{\alpha\beta\gamma\delta}=\left(  \frac
{\varkappa^{4}}{54}+\frac{\dot{\varkappa}^{2}}{3}+\frac{\varkappa^{2}%
\dot{\varkappa}}{9}\right)  +\frac{\left(  4\dot{\varkappa}+\varkappa
^{2}\right)  }{12\gamma}\bar{\lambda}^{2}-\frac{\varkappa}{\gamma^{\frac{3}%
{2}}}\lambda_{1}\lambda_{2}\lambda_{3}+\frac{3}{16\gamma^{2}}\bar{\lambda}^{4}
\label{36ffff}%
\end{equation}
which are quoted above, for convenience of \ the reader, we find
\begin{equation}
R=-\frac{8\varepsilon_{m}}{\left(  1+3\varepsilon_{m}t^{2}\right)  ^{2}%
},\text{ \ }R_{\alpha\beta}R^{\alpha\beta}=\frac{28\varepsilon_{m}^{2}%
}{\left(  3\varepsilon_{m}t^{2}+1\right)  ^{4}}, \label{36a}%
\end{equation}
and%

\begin{equation}
R_{\alpha\beta\gamma\delta}R^{\alpha\beta\gamma\delta}=\frac{8\varepsilon
_{m}^{2}\left(  12t^{4}\varepsilon_{m}^{2}+6t^{2}\varepsilon_{m}+5\right)
}{\left(  3\varepsilon_{m}t^{2}+1\right)  ^{4}}-\frac{16}{t^{4}}\left(
\frac{3\varepsilon_{m}t^{2}}{1+3\varepsilon_{m}t^{2}}\right)  ^{5/2}\left(
p_{1}p_{2}p_{3}+\frac{2}{27}\right)  . \label{36b}%
\end{equation}
Hence, not only the invariants but also all their derivatives are finite at
any moment of time. For $\varepsilon_{m}t^{2}\gg1$ the solution approaches the
Kasner solution in the leading order, up to higher order corrections. For
instance, at large $t$ the scalar curvature is not exactly vanishing and it is
of order%
\begin{equation}
R\sim\frac{1}{\varepsilon_{m}}R_{\alpha\beta\gamma\delta}R^{\alpha\beta
\gamma\delta}, \label{36c}%
\end{equation}
which in $\varepsilon_{m}^{-1}$ times exceeds the vacuum polarization
corrections, and at low curvatures are negligible, unless $\varepsilon_{m}$ is
not much smaller than the Planckian value.

Finally we discuss how the presence of matter can influence the behavior of
Kasner solutions in the vicinity of a bounce. It is clear that if at some
moment in the past at $t_{0}<-\varepsilon_{m}^{-1/2}$ the curvature
contribution in $\left(  \ref{20a}\right)  $ dominates over the the usual
matter, that is,
\[
\frac{\bar{\lambda}^{2}}{\gamma\left(  t_{0}\right)  }>T_{0}^{0}\left(
t_{0}\right)  .
\]
then $T_{0}^{0}\left(  t\right)  $ in $\left(  \ref{20a}\right)  $ can be
neglected if the matter equation of state satisfies $w<1.$ In fact, $T_{0}%
^{0}\propto\gamma^{-\frac{1+w}{2}}$ does not grow as fast as $\bar{\lambda
}^{2}/\gamma$ if $\gamma$ decreases, and therefore can be neglected when we
approach the bounce. On the other hand when $w>1$ the matter term can start to
dominate well before the bounce, and in this case the bounce happens as in the
Friedmann universe considered above.

If matter (for instance, the scalar field) changes the equation of state from
$w\gg1$ before the bounce to $w<1$ after the bounce, the Kasner universe which
was strongly anisotropic can create homogeneous isotropic Friedmann universe
after the bounce. In fact, at large $\gamma$ in contracting Kasner universe,
the usual matter with $w\gg1$ can be completely ignored because $T_{0}%
^{0}\propto\gamma^{-\frac{1+w}{2}}$ and this matter becomes relevant only at
small $\gamma$ close to the bounce. If the equation of state during or after
the bounce changes to $w<1$ then during the expansion stage, $T_{0}^{0}$ will
be decaying not as fast as $\bar{\lambda}^{2}/\gamma,$ and finally begins to
dominate. In this case $\dot{\gamma}/\gamma\propto\gamma^{-\frac{1+w}{4}}$ and
the second term in $\left(  \ref{32}\right)  ,$ which spoils the isotropy and
decay as $\lambda_{i}/\sqrt{\gamma}$ will finally become negligible compared
to $\dot{\gamma}/\gamma.$ As a result the universe will become isotropic. This
opens the possibility to solve the isotropy problem in bouncing cosmology.

\section{Discussion}

We have shown above that the singularities in cosmological solutions can be
easily removed by making the longitudinal degree of freedom of gravity to be
dynamical. The presence of mimetic field needed to generate Born-Infeld
corrections to the Einstein theory adds extra \textquotedblleft dust
like\textquotedblright\ degree of freedom to gravity, which can serve as
mimetic dark matter. It is rather curious that the non-singluar modification
of General Relativity delivers for \textit{free} the realistic candidate for
Dark Matter. Besides of the appearance of mimetic matter, Einstein's equations
are significantly modified only at very high curvatures which are close to the
limiting one.

In the considerations above we have chosen a particular function $f\left(
\square\phi\right)  $ to get the simplest possible equations$.$ One can wonder
to what extent this choice is unique? Our preliminary investigation shows that
there is a whole class of functions $f$ which would lead to theories with
limiting curvature. However this class is rather restricted. In particular,
the necessary ingredients common to all these functions are 1) they have to
contain a Born-Infeld type term, and 2) the derivative $\frac{df}{d\chi}$ must
remain finite when $\chi$ approaches its limiting value $\chi_{m}.$

The non-singluar models we have constructed can be rather important for
bouncing cosmologies. In fact, in these models a bounce happens within very
short time interval of order $t\sim\varepsilon_{m}^{-1/2}$ and if the limiting
curvature would be of the order of Planck value, this time would be the
Planckian time. Outside this time interval Einstein's theory is well
applicable. Using causality one can argue that in this case perturbations
generated in a contracting universe on the supercurvature scale can be
re-translated to an expanding universe practically without any change.
However, this question requires further quantitative investigation.

The idea of limiting curvature can also have rather severe consequences for
inflationary universe. First of all, notice that if we want to use the stage
of accelerated expansion for amplifying quantum fluctuations, observed in
numerous CMB experiments, the limiting curvature cannot be below the
inflationary scale, that is, it cannot be less than the Planckian scale more
than just by few orders of magnitude. When the limiting curvature is below the
self-reproduction scale the multiverse is avoided. The flat inflationary
potentials, favoured by observations, also look more natural from the point of
view of the idea of limiting curvature.

The theory we have considered here is a classical theory and if the limiting
curvature is well below the Planck scale we can safely ignore quantum
corrections. However, any field gravitates, and therefore the highly energetic
quanta which in Einstein's theory would produce the curvature exceeding the
limiting one, must be either prohibited or modified. This is why we expect
that the natural cut-off in quantum field theories, which is normally taken to
be of the order of Planck value, can be well below the Planck scale in
theories with limiting curvature.

The considerations above are restricted to highly symmetric space-times. One
could naturally address the question whether the curvature will remain bounded
generically for arbitrary inhomogeneous spaces. In an accompanying paper
\cite{BH} we show that the singularity is also avoided in the case of a black
hole and give the explicit solution for it. Moreover analyzing the expressions
for the curvature invariants one \ could argue that the limiting curvature is
the generic property of arbitrary space-time irrespective of their spatial
curvature. However these questions require a more serious investigation.

\textbf{{\large {Acknowledgments}}}

The work of A. H. C is supported in part by the National Science Foundation
Grant No. Phys-1518371. The work of V.M. is supported in part by Simons
Foundation grant 403033TRR 33 \ and \textquotedblleft The Dark
Universe\textquotedblright\ and the Cluster of Excellence EXC 153
\textquotedblleft Origin and Structure of the Universe\textquotedblright%
.\bigskip\

\end{document}